\documentstyle[aps]{revtex}
\newcommand{\ds}{\displaystyle}
\newcommand{\dsf}{\ds\frac}

\newcommand{\beq}{\begin{equation}}
\newcommand{\eeq}{\end{equation}}

\large

\begin{document}
\large
\begin{center}
\Large\bf
On the stability of the critical state with inhomogeneous
temperature in composite superconductors
\vskip 0.1cm
{\normalsize\bf N.A.\,Taylanov}\\
\vskip 0.1cm
{\large\em Theoretical Physics Department,\\
Scientific Research Institute for Applied Physics,\\
National University of Uzbekistan,\\
Tashkent, 700174, Uzbekistan\\
E-mail: taylanov@iaph.tkt.uz}
\end{center}
\begin{center}
{\bf Abstract}
\end{center}
\begin{center}
\mbox{\parbox{14cm}{\small

        The problem of the thermal and magnetic destruction of the critical state
in composite superconductors is investigated. The initial distributions of
temperature and electromagnetic field are assumed to be essentially
inhomogeneous. The limit of the thermomagnetic instability in
quasi-stationary approximation is determined. The obtained integral criterion,
unlike the analogous criterion for a homogeneous temperature profile,
is shown to take into account the influence of any part of the superconductor
on the threshold for critical-state instability.
}}
\end{center}
\vskip 3mm

While dealing with instabilities of the critical state in hard
superconductors, the character of the temperature distribution $T(x,t)$
and that of the electromagnetic field $\vec E(x,t)$ are of substantial
practical interest [1]. This derives from the fact that thermal and magnetic
distractions of the critical state caused by Joule self-heating are
defined by the initial temperature and electromagnetic field distributions.
Hence, the form of the temperature profile may noticeably influence the
criteria of critical-state stability with respect to jumps in the magnetic
flux in a superconductor. Earlier (cf., e.g., [2]), in dealing with
this problem, it was usually assumed that the spatial distribution of
temperature and field were either homogeneous or slightly inhomogeneous.
However, in reality, physical parameters of superconductors may be
inhomogeneous along the sample as well as in its cross-sectional plane.
Such inhomogeneities can appear due to different physical reasons.
First, the vortex structure pinning can be inhomogeneous due to the
existence of weak bonds in the superconductor. Second, inhomogenety
of the properties may be caused by their dependence on the magnetic
field $H$. Indeed, the field $H$ influences many physical quantities,
such as the critical current density $j_c$, the differential conductivity
$\sigma_d$, and the heat conductivity $k$.

In the present paper, the temperature distribution in the critical state
of composite superconductors is investigated in the quasi-stationary
approximation.
It is shown that the temperature profile can be essentially inhomogeneous,
which affects the conditions of initiation of magnetic flux jumps.

The evolution of thermal ($T$) and electromagnetic
($\vec E, \vec H$) perturbations in superconductors is described by a
nonlinear heat conduction equation [3],
\beq
\nu\dsf{dT}{dt}=\nabla [\kappa\nabla T]+\vec j\vec E,
\eeq
a system of Maxwell's equations,
\beq
rot\vec E=-\dsf{1}{c}\dsf{d\vec H}{dt},
\eeq
\beq
rot{\vec H}=\dsf{4\pi}{c}\vec j
\eeq
and a critical-state equation
\beq
\vec j=\vec j_{c}(T,\vec H)+\vec j_{r}(\vec E).
\eeq
Here $\nu=\nu(T)$ is the specific heat, $\kappa=\kappa(T)$ is the thermal
conductivity respectively; $\vec j_c$ is the critical current density and
$\vec j_r$ is the active current density.

We use the Bean-London critical state model to
describe the $j_{c}(T,H)$ dependence, according to which
$j_{c}(T)=j_0-a(T-T_{0})$ [4], where the parameter $a$ characterizes
thermally activated weakening of Abrikosov vortex pinning on crystal
lattice defects, $j_{0}$ is the equilibrium current density, and
$T_0$ is the temperature of the superconductor.

The dependence $j_r(E)$ in the region of sufficiently strong electric
fields ($E\ge E_f$; where $E_f$ is the limit of the linear region of the
current-voltage characteristic of the sample [2]) can be approximated
by a piecewise-linear function $j_r\approx\sigma_f E$, where
$\sigma_f=\dsf{\eta c^2}{H\Phi_0}\approx \dsf{\sigma_n H_{c_2}}{H}$ is the
effective conductivity in the flux flow regime and $\eta$ is the viscous
coefficient, $\Phi_0=\dsf{\pi h c}{2e}$ is the magnetic flux quantum,
$\sigma_n$ is the conductivity in the normal state, $H_{c_2}$ is the upper
critical magnetic field. In the region of the weak fields $(E\le E_f)$, the
function $j_r (E)$ is
nonlinear. This nonlinearity is associated with thermally activated
creep of the magnetic flux [5].

Let us consider a superconducting sample placed into an external magnetic
field $\vec H=(0, 0, H_{e})$ increasing at a constant rate
$\dsf{d\vec H}{dt}=\dot H$=const. According to the Maxwell equation (2),
a vortex electric field $\vec E=(0, E_e, 0)$ is present. Here $H_e$ is
the magnitude of the external magnetic field and $E_e$ is the magnitude
of the back-ground electric field. In accordance with the concept of
the critical state, the current density and the electric field must be
parallel: $\vec E\parallel \vec j$; where $H_e$ is the
amplitude of the external magnetic field and $E_e$ is the amplitude of the
external electric field.

       The thermal and electromagnetic boundary conditions for the
Eqs. (1)-(4) have the form
\beq
\begin{array}{ll}
\left.\kappa \dsf{dT}{dx}\right|_{x=0}+w_0[T(0)-T_0]=0\,,\qquad &
T(L)=T_{0}\,,\\
\quad \\
\left.\dsf{dE}{dx}\right|_{x=0}=0\,, &   E(L)=0\,,
\end{array}
\eeq

For the plane geometry (Fig.1.) and for the boundary conditions
$H(0)=H_e,\quad H(L)=0$, the magnetic field distribution is
$H(x)=H_e(L-x)$,
where $L=\dsf{cH_e}{4\pi j_c}$ is the depth of magnetic flux penetration
into the sample and $w_0$ is the coefficient of heat transfer to the
cooler at the equilibrium temperature $T_0$.

The condition of applicability of Eqs. (1)-(4) to the description of
the dynamics of evolution of thermomagnetic perturbations are discussed
at length in [1].

In the quasi-stationary approximation, terms with time derivatives can
be neglected in  Eqs. (1)-(4). This means that the heat transfer from
the sample surface compensates the energy dissipation arising in the
viscous flow of magnetic flux in the medium with an effective conductivity
$\sigma_f$.
In this approximation, the solution to Eq. (2) has the form

\beq
E=\dsf{\dot H}{c}(L-x).
\eeq

Upon substituting this expression into Eq. (1) we get an inhomogeneous
equation for the temperature distribution  $T(x,t)$,
\beq
\dsf{d^2\Theta}{d\rho^2}-\rho\Theta=f(\rho).
\eeq
Here we introduced the following dimensionless variables
$$
f(\rho)=-[1+r\omega\rho]\dsf{j_0}{aT_0},\\
\qquad
\Theta=\dsf{T-T_0}{T_0},\\
\qquad
\rho=\dsf{L-x}{r},\\
$$
and the dimensionless parameters
$\omega =\dsf{\sigma_f \dot H}{cj_0}$,
and, $r=\left(\dsf{c\kappa}{a \dot H L^2} \right)^{1/3}$,
where $r$ characterizes the spatial scale of the temperature profile
inhomogeneity in the sample.
Solutions to Eq. (7) are Airy functions, which can be expressed
through Bessel functions of the order 1/3 [6]

\beq
\Theta (\rho) = C_1\rho^{1/2}K_{1/3}\left(\dsf{2}{3} \rho^{3/2}\right)
+ C_2\rho^{1/2}I_{1/3}\left(\dsf{2}{3} \rho^{3/2}\right)+\Theta_0(\rho),
\eeq
$$
\Theta_0(\rho) = \rho^{1/2}K_{1/3}\left(\dsf{2}{3}\rho^{3/2}\right)
\int_{0}^{\rho}
[1+r\omega \rho_1]\rho_{1}^{3/2}I_{1/3}
\left(\dsf{2}{3}\rho_{1}^{3/2}\right)d\rho_1 -
$$
$$
\rho^{1/2}I_{1/3}\left(\dsf{2}{3}\rho^{3/2}\right)\int_{0}^{\rho}
[1+r\omega \rho_1]\rho_{1}^{3/2}K_{1/3}
\left(\dsf{2}{3}\rho_{1}^{3/2}\right)d\rho_1,
$$
where $C_1$ and $C_2$ are integration constants, which are determined by
the boundary conditions to be
$$C_1=0,\quad C_2=\dsf{-w_0L\Theta(0)+\left.\kappa\dsf{d\Theta}{d\rho}
\right|_{\rho=\dsf{L}{r}}}
{\left.\left[w_0\left(\dsf{L}{r}\right)^{1/2}I_{1/3}
\left(\dsf{2}{3}\rho^{-3/2}\right)-
2\dsf{d}{d\rho}
\left(\rho^{1/2}I_{1/3}\left(\dsf{2}{3}\rho^{3/2}\right)\right)
\right]\right|_{\rho=\dsf{L}{r}}}
$$
From the Maxwell equation (2), the temperature inhomogeneity parameter can
be expressed in the form
\beq
\alpha=\dsf{r}{L}=\left[\dsf{4\pi\nu j_0}{aH_{e}^{2}}
\dsf{H_e}{\dot Ht_\kappa}\right]^{1/3}
\eeq
        It is evident that $\alpha\sim 1$ near the threshold for a flux jump,
when $\dsf{aH_{e}^{2}}{4\pi\nu j_0}\sim 1$, even under the quasi-stationary
heating condition $\dsf{\dot H t_{\kappa}}{H_e}<<1$; where
$t_{\kappa}=\dsf{\nu L^2}{\kappa}$ is the
characteristic time of the heat conduction problem.

       Let us estimate the maximum heating temperature $\Theta_m$
in the isothermal case $w=\dsf{\kappa}{L}\ge 1$.
The solution to Eq. (7) can be represented in the form
\beq
\Theta(x)=\Theta_m-\rho_0\dsf{(x-x_m)^2}{2},
\eeq
near the point at which the temperature is a maximum, $x=x_m$  (Fig.1).

With solution (10) being approximated near the point $x_m=\dsf{L}{2}$
with the help of the thermal boundary conditions, the coefficient
$\rho_0$ can be easily determined to be $\left(\dsf{8}{L^2}\right)\Theta_m$
and the temperature can be written as
\beq
\Theta(x)=\Theta_m\left[1-\dsf{4}{L^2}\left(x-\dsf{L}{2}\right)^2\right],
\eeq
Substituting this solution into Eq.(7), the superconductor maximum
heating temperature due to magnetic flux jumps can be estimated as
\beq
\Theta_m=\dsf{\left[j_0+\dsf{\sigma_f \dot H}{c}(L-x_m)\right]\dsf{\dot H}{c \kappa T_0}
(L-x_m)}{\dsf{\gamma}{L^2}+\dsf{a\dot H}{c\kappa}(L-x_m)}.
\eeq
For a typical situation, when
$\dsf{\gamma}{L^2}<<\dsf{a\dot H}{c\kappa}(L-x_m)$
the estimation for $\Theta_m$ is
\beq
\Theta_m\approx\left[j_0+\dsf{\sigma_f \dot H}{c}(L-x_m)\right]\dsf{\dot HL^2}
{c\kappa T_0}(L-x_m).
\eeq
Here, the parameter $\gamma\sim 1$ (for a parabolic temperature profile
$\gamma\sim 8$). It is easy to verify that for typical values of
$j_0=10^6 A/cm^2$,
$\dot H=10^4$ $G/s$, and $L=0,01$ cm the heating is sufficiently low:
$\Theta_m<<1$.
In the case of poor sample cooling, $w=1\div 10 erg/(cm^2 s K)$,
the $\Theta_m$ is
$$
\Theta_m=\dsf{\dot Hj_0L^2}{cw_0T_0}\approx 0,5;
$$

i.e., the heating temperature can be as high as
$\delta T_m=T_0\Theta_m\sim 2 K$.
One can see that in the case of poor sample cooling, the heating can be
rather noticeable and influences the conditions of the thermomagnetic
instability of the critical state in the superconductor.

        Let us investigate the stability of the critical state with respect
to small thermal $\delta T$ and electromagnetic $\delta E$ fluctuations
in the quasi-stationary approximation. We represent solutions to Eqs. (1)-(4)
in the form
\beq
\begin{array}{l}
T(x,t)= T(x)+\exp\left\{\ds\dsf{\lambda t}{t_{\kappa}}\right\}
\delta T\left(\dsf{x}{L}\right)\,,\\
\quad \\
E(x,t)= E(x)+\exp\left\{\ds\dsf{\lambda t}{t_{\kappa}}\right\}
\delta E\left(\dsf{x}{L}\right)\,,
\end{array}
\eeq
where $T(x)$ and $E(x)$ are solutions to the unperturbed equations obtained
in the quasi-stationary approximation describing the background distributions
of temperature and electric field in the sample and
$\lambda$ is a parameter to be determined. The instability region is
determined by the condition that $Re{\lambda}\ge 0$.
From solution (14), one can see that the characteristic time of thermal and
electromagnetic perturbations $t_j$ is of the order of
$t_\kappa/\lambda$.
Linearizing Eqs. (1)-(4) for small perturbations
$\left(\ds\dsf{\delta T}{T(x)}, \dsf{\delta E}{E(x)}<<1\right)$ we obtain the
following equations in the quasi-stationary approximation:
\beq
\begin{array}{l}
\nu\dsf{\lambda}{t_\kappa}\delta T=\dsf{\kappa}{L^2}\dsf{d^2\delta T}{dx^2}
+\left[j(x)+\sigma_{f}E(x)\right]\delta E-aE(x)\delta T\,,\\
\quad\\
\dsf{1}{L^2}\dsf{d^2\delta E}{dx^2}=\dsf{4\pi\lambda}{c^2t_\kappa}
\left[\sigma_f\delta E - a\delta T\right]\,.
\end{array}
\eeq

Eliminating the variable $\delta T$ between Eqs. (15), we obtain a fourth-
order differential equation with variable coefficients for the
electromagnetic field  $\delta E$:

\beq
\dsf{d^4\delta E}{dz^4}-
\left[\lambda(1+\tau)+\dsf{E(z)}{E_\kappa}\right]\dsf{d^2\delta E}{dz^2}+
\lambda \left[\lambda\tau-B(z)\right]\delta E=0\,.
\eeq

Here, we introduced the following dimensionless variables:
$z=\dsf{x}{L},
\quad B(z)=\dsf{4\pi aL^2}{c^2\nu} j(z),
\quad j(z)=\sigma_f E(z)-[j_0-a(T(z)-T_0)],
\quad E(z)=\dsf{\dot H L}{c}(1-z),
\quad \tau=\dsf{4\pi \sigma_f\kappa}{c^2\nu},
\quad E_{\kappa}=\dsf{\kappa}{aL^2}.
\quad \nu=\nu_0\left(\dsf{T}{T_0}\right)^3,
\quad \kappa=\kappa_0\left(\dsf{T}{T_0}\right)$.

        One should keep in mind that the variable $T(z)$ and $E(z)$
are given by Eq. (8), in which $\rho=\dsf{L}{r}(1-z)$.
Using the relation between $\delta T$ and $\delta E$ given by Eqs. (15),
we write the boundary conditions to Eq. (16) in the form:

\beq
\begin{array}{l}
\quad\left.\dsf{d^2\delta E}{dz^2}\right|_{z=1}=0,
\quad\left.\dsf{d^3\delta E}{dz^3}\right|_{z=0}=
-W\left.\left[\dsf{d^2\delta E}{dz^2}-\lambda\tau\delta E\right]
\right|_{z=0}\,,\\
\quad\\
\left.\delta E\right|_{z=1}=0,
\quad\left.\dsf{d\delta E}{dz}\right|_{z=0}=0\,.
\end{array}
\eeq

where $W=\dsf{w_0 L}{\kappa}$ is the dimensionless thermal impedance.

        The condition for the existence of a nontrivial solution to Eq. (16)
subject to boundary conditions (17) allows one to determine the boundary
of the critical-state thermomagnetic instability in a superconducting sample.
This problem is complicated, and its analytical solution cannot be found in a
closed form. We will consider the development of thermomagnetic instability
in the dynamical approximation, which is valid for composite superconductors
with high value of $\sigma_f E$.

The dynamical character of the instability
development leads to the predominance of heat diffusion over
magnetic flux diffusion in the sample: $\tau=\dsf{D_t}{D_m}>>1$ [1], where
$D_t=\dsf{\kappa}{\nu}$ and $D_m=\dsf{c^2}{4\pi \sigma_f}$ are the
coefficients of the thermal and magnetic diffusion, respectively.

        In this case, as seen from Eq. (14), the characteristic
times $t_j$ of temperature
and electromagnetic field perturbations have to satisfy the
inequalities $t_j>>t_\kappa$ $(\lambda<<1)$ and
$t_j<<t_m$ ($\lambda\tau>>1$), where $t_\kappa=\dsf{L^2}{D_t}$
and $t_m=\dsf{L^2}{D_m}$ are the characteristic times of the thermal and
magnetic diffusion, respectively.

As well known that [1], the effective magnitude of
conductivity $\sigma_f$ is greater in composite superconductors
than the one in hard superconductors.
We can assume that induced normal current $\sigma_f E$ compensates the
decreasing of critical current $j_c(T)$, caused by increasing of temperature
and obviously prevents magnetic flux penetration into the sample. In this
case we can neglect the moving of magnetic flux.  In the other word,
thermomgnetic instability developes slowly as compared with the heat
diffusion with characteristic time of increasing
$t_j\sim\dsf{t_\kappa}{\lambda}>>t_\kappa$
($\lambda<<1)$ or $t_m>>\dsf{t_m}{\lambda\tau}= \dsf{t_\kappa}{\lambda}=t_j$
$(\lambda\tau>>1$).

In the approximation  ($\tau>>1$, $\lambda\tau>>1$, $\lambda<<1$)
Eq. (16) is reduced to a lower order differential equation

\beq
\dsf{d^2\delta E}{dz^2}+\left[\lambda-\dsf{B(z)}{\tau}\right]\delta E=0\,.
\eeq

       In the case of $\tau>>1$, the instability threshold depends
on the electrodynamic boundary conditions at the surface of the sample only
slightly. Therefore, the electrodynamic boundary conditions at the boundaries
of the current-carrying layer $(z=0, z=1)$ can be neglected and one can
keep only the thermal boundary conditions to Eq. (18).

Multiplying Eq. (18) by $\delta E$ and integrating the result with respect
to $z$ over the interval $0<z<1$, we obtain

\beq
\lambda=
\dsf{\dsf{1}{\tau}\ds\int\limits_{0}^{1}
B(z)\ds\delta E^2dz-
\ds\int\limits_{0}^{1}\left(\dsf{d\delta E}{dz}\right)^2dz}
{\ds\int\limits_{0}^{1}
\delta E^2dz}
\eeq
where we use the equality

$$
\ds\int\limits_{0}^{1}\dsf{d^2\delta E}{dz^2}\delta E dz
=\delta E(z)\left(\dsf{d\delta E}{dz}\right)-
\ds\int\limits_{0}^{1}\left(\dsf{\delta E}{dz}\right)^2dz=
-\ds\int\limits_{0}^{1}\left(\dsf{\delta E}{dz}\right)^2dz
$$
and the boundary conditions. The right-hand side of Eq. (19) has a
minimum at $\lambda = \lambda_c$:

\beq
\lambda_{c}=
\dsf{\dsf{1}{\tau}\ds\int\limits_{0}^{1}B(z)\delta E^2dz}
{\ds\int\limits_{0}^{1}\delta E^2dz}\,,
\eeq
then
\beq
\dsf{1}{\tau}\ds\int\limits_{0}^{1}B(z)n_E^2 dz=
\dsf{\ds\int\limits_{0}^{1}\left(\dsf{d\delta E}{dz}\right)^2dz}
{\ds\int\limits_{0}^{1}\delta E^2dz}.
\eeq
Here we introduced the following unit vector
$$
n_E^2=\dsf{\delta E^2}{\ds\int\limits_{0}^{1}\delta E^2dz}\,, \qquad
\ds\int\limits_{0}^{1}n_ E^2dz=1.
$$

Since we do not know the function $\delta E(z)$, we try, following [7],
to obtain an integral estimation of the instability growth increment and
the low
boundary of its occurrence. The behavior of the integrand in Eq. (21) is
basically determined by the factor
$E=\dsf{\dot H L}{c}(1-z)$, which is equal to zero at $z=1$ (the other
factors
change more smoothly). Hence, the integrand reaches its maximum at
$z=0$ and the upper estimate for $\lambda_c$ is

\beq
\lambda_c\le
\dsf{B(0)}{\tau}\,.
\eeq

        It is evident that $\lambda_c<<1$ and $\lambda_c\tau>>1$ at
$\tau>>1$. Numerical evaluation gives $\lambda_c \approx 0.1\div 10^{-2}$
at $\tau=10^{3}$.

        Equations (21) and (22) enable one to write the instability
occurrence criterion in the form

\beq
\dsf{1}{\tau}\ds\int\limits_{0}^{1}B(z)
n_E^2dz>\dsf{\ds\int\limits_{0}^{1}\left(\dsf{d\delta E}{dz}\right)^2}
{\ds\int\limits_{0}^{1}\delta E^2dz}\,.
\eeq

        This criterion essentially depends on the boundary conditions and
the functions $j(z)$, $E(z)$, $T(z)$.
Figure 2 presents graph of the function $T(z)$.
Inequality (23) can be strengthened by means of an evaluation,

\beq
\ds\int\limits_{0}^{1}
\ds\left(\dsf{d\delta E}{dz}\right)^2dz>\dsf{\pi^2}{4}\ds\int\limits_{0}^{1}
\delta E^2dz
\eeq
which can be easily verified by expanding the function $\delta E(z)$
in a Fourier series:

$$
\delta E(z)=A_m cos{\dsf{\pi z(2m+1)}{2}}\,.
$$

Let us now try to strengthen inequality (23) further. For this purpose,
we consider the integral
$$
\int\limits_{0}^{1}g(z)(n_{E}^{2}-1)dz=
\int\limits_{0}^{z_1}g(z)(n_{E}^{2}-1)dz+
\int\limits_{z_1}^{0}g(z)(n_{E}^{2}-1)dz.
$$
The last term can be represented in the form
$$
\int\limits_{0}^{1}g(z)(n_{E}^{2}-1)dz=
(g_+-g_-)\int\limits_{0}^{1}g(z)(n_{E}^{2}-1)dz,
$$
taking intermediate values of the $g_-$ in the range $z<z_1$ and
$g>g_-$ in the range $z_1<z<1$ outside the integral. It is evident
(Fig.3) that
$$
\int\limits_{0}^{1}g(z)(n_{E}^{2}-1)dz\le 0
$$
or
\beq
\int\limits_{0}^{1}g(z)n_{E}^{2}dz\le \int\limits_{0}^{1}g(z)dz.
\eeq

With inequality (25), the instability occurrence criterion can be
represented in the form

\beq
\ds\int\limits_{0}^{1}B(z)dz\ge\dsf{\pi^2}{4}\tau\,.
\eeq

        Inequality (26), unlike the analogous criterion for a homogeneous
temperature profile, has an integral character and takes into account the
influence of each part of the superconductor on the threshold for the
superconducting-state instability. If condition (26) is satisfied, then
small fluctuations of temperature $\delta T$ and electric field $\delta E$
in the superconductor will exponentially increase with time. The most probable
result of the development of such an instability would be a transition
from a critical state to a resistive one.

        We should emphasis that this result was obtained
for an arbitrary temperature dependence of thermophysical parameters
$\nu$ and $\kappa$ of superconducting material and for an arbitrary
function $j(H)$. Moreover, since the system of equations (1)-(4)
is invariant with respect to an arbitrary translation, the wave propagation
condition can be found for an arbitrary critical current density dependence
on T and H.

\newpage

\centerline{\large\bf FIGURE}

Fig.1. The distrubation of the temperature profile $\Theta (x)$.

Fig.2. Plots of the function $T(z)$.

Fig.3. Plots of the function $g(z)$.

\newpage
\centerline{\large \bf NIZAM A.TAYLANOV, Golibjon R. Berdiyorov}
\begin{tabbing}
{\large \bf Address:} \\
Theoretical Physics Department and\\
Institute of Applied Physics,\\
National University of Uzbekistan,\\
Vuzgorodok, 700174, Tashkent, Uzbekistan\\
Telephone:(9-98712),461-573, 460-867.\\
fax: (9-98712) 463-262,(9-98712) 461-540,(9-9871) 144-77-28\\
e-mail: taylanov@iaph.tkt.uz\\

\end{tabbing}

\end{document}